\documentstyle[aps,prl,twocolumn,epsf]{revtex} 
\begin{document} 

\draft

\title{Tracer Dispersion in a Self-Organized Critical System}

\author{Kim Christensen,$^{{\ast}{\dag}}$ \'{A}lvaro Corral,$^{\ddag}$
Vidar Frette,$^{{\ast}{\S}}$ Jens Feder,$^{\ast}$
and Torstein J{\o}ssang$^{\ast}$ 
}
\address{
$^{\ast}$Department of Physics, University of Oslo, \\
P.O. Box 1048, Blindern, N-0316 Oslo 3, \\
Norway \\ 
$^{\dag}$Present address: Instituto de F\'{\i}sica, Universidade Federal do Rio
de Janeiro, UFRJ \\
Cxp. 68528, 21945-970 Rio de Janeiro, RJ, \\
Brazil \\ 
$^{\ddag}$Departament de F\'{\i}sica Fonamental, Universitat de Barcelona, \\
Diagonal 647, E-08028 Barcelona, \\
Spain \\ 
$^{\S}$Present address: Department of Physics of Complex Systems, \\
The Weizmann Institute of Science, Rehovot 76100, \\
Israel}

\date{\today}

\maketitle

%\newpage
%\begin{center}{\bf Abstract}\end{center}
\begin{abstract} 
We have studied experimentally transport properties 
in a slowly driven granular system which recently
was shown to display self-organized criticality
[Frette {\em et al., Nature} {\bf 379}, 49 (1996)].
Tracer particles were added to a pile and their
transit times measured. The distribution of transit
times is a constant with a crossover to a decaying
power law. The average transport velocity decreases with system size.
This is due to an increase in the active zone depth with system size.
The relaxation processes generate coherently moving regions of
grains mixed with convection.
This picture is supported by considering transport
in a $1D$ cellular automaton modeling the experiment. 
\end{abstract} 

%\vspace{0.5cm}
%\noindent 

%\begin{multicols}{2}
%\newpage
The avalanches that occur when grains are dropped onto a pile
illustrate the spontaneous generation of complexity
in simple dynamical systems \cite{BTW87}.
When grains are dropped onto a finite base, a pile
builds up. However, it cannot become infinitely high, and,
eventually, the system settles
in a stationary state where the outflux over the edge of the
base on average
equals the influx.
Intermittent flow of grains down the
slope of the pile (small and large avalanches)
maintain the system in this state.
Bak, Tang, and Wiesenfeld constructed a $2D$ cellular automaton
of a slowly driven dynamical system. They showed, that the ``pile''
spontaneously evolves, or self-organizes, into a state with
avalanches of all sizes distributed according to a power law,
that is, there is no internal system-specific scale.
Because of the lack of any characteristic avalanche size,
the system is referred to as critical \cite{BTW87}. 

It has been a longstanding question whether real granular
systems display self-organized criticality (SOC)
when slowly driven.
Recently, however, an experiment on a quasi one-dimensional
pile of rice has 
shown that the occurrence of SOC depends on
details in the grain-level dissipation
mechanisms \cite{nature95}.
With nearly spherical grains, a characteristic avalanche size appeared,
inconsistent with SOC. Only with sufficiently elongated grains,
avalanches with a power-law distribution occurred.
We focus in this Letter on the transport that results from
avalanches in the system displaying SOC.
The elongated rice grains could pack in a variety of ways,
and each avalanche replaced, locally or globally,
one surface configuration with another. Thus a dynamically
varying medium disorder (coupled to the relaxation processes)
was generated. This is conceptually different
from transport in media with a quenched disorder,
see e.g. Refs. \cite{HB87,BG90}.
Furthermore, in SOC systems, a small perturbation may lead to arbitrarily
large avalanches, and it is not clear at all, how this affects
the transport properties.
Thus it is quite surprising, that there are no
experiments and only a few
theoretical and numerical studies 
on transport in systems displaying SOC \cite{HK89,CCGS90,BJ92}. 

We have measured the transit times of colored tracer particles
in the rice pile.
Experimentally, we find that the distribution
of transit times 
is essentially constant for small transit times $T$
%(when displayed in a log-log plot)
and decays as a power law
for large $T$. A finite-size scaling analysis shows, that
the average transit time $\langle T \rangle \propto L^{\nu}$ where
$L$ is the system size and $\nu = 1.5 \pm 0.2$. Thus the
average velocity of tracer particles
$\langle V \rangle \propto L/\langle T \rangle$
decreases with system size. This implies the existence of
long-range correlations in the rice pile.
Correlations in the sequence of transit times decay exponentially
with a correlation time that increases with system size but have a
crossover to a power-law decay.
This indicates that the system has coherently moving regions, the
size of which increases with system size.
A $1D$ cellular automaton model, related to the experiment,
displays SOC behavior and the distribution of 
transit times agrees qualitatively with the experimental findings.
The scaling of the average transit time
with system size is reproduced and hence the decrease of the
average velocity with increasing system size.
This phenomenon is related to an increase in the active
zone depth as the system size is increased.  
Finally, the numerical and experimental results for the correlations in
the transit times are very similar.

The experimental system consisted of
a rice pile confined between two
$5\,$mm thick glass plates supported by $15\,$mm thick
$100\,$cm $\!\times\!$ $120\,$cm
polymethylmethacrylate plates.
Aluminum rods were inserted between the glass plates to form
a vertical wall at one side and a variable base with length $\cal L$ 
of the quasi one-dimensional pile.
The other side was open, allowing grains to fall
off the pile.
Grains of ``Geisha Naturris'' from Stabburet a.s.
(Troll{\aa}sen, Norway) with length $\delta = 7.6 \pm 0.9\,$mm and
width $2.0 \pm 0.1\,$mm were slowly fed into the gap between the
plates close to the vertical wall using an \"{O}vrum Nibex
500 (\"{O}vrum, Sweden) single seed machine.
We used a plate separation of $d = 6\,$mm and system
sizes ${\cal L} = 15, 30, 60,$ and $85.7\,$cm, which,
expressed in units of the grain length $\delta$, are
${\cal L}/\delta = L = 20, 39, 79,$ and
$113.$
The injection rate was 2--3 uncolored grains every $7.7\,s$ or, on average,
20 grains/min.
When the pile had reached the stationary state,
color coded particles were added to the pile by hand and the
transit times measured. 
A total 
number of $400$ tracer particles for the two smaller systems
and $800$ for the two larger systems were used.
Except for the smallest system, where
a color coded tracer particle was injected every 2nd minute,
the rate was 1 tracer particle every 4th minute.
The injection of uncolored grains was continued until all tracers had left
the system.

Figure\ 1 shows a part of the $L = 113$ system at
a late stage of the experiment.
Note that some of the tracer particles are 
close to the surface layer while others are
buried quite deep in the pile.
These tracers had to be released
by large avalanches in order to move on.
However, large avalanches occurred with
a very small probability, and hence deeply buried particles tended
to stay in the system for a very long time.
Figure\ 2 is a full record of the experimental findings for the
$L = 113$ rice pile. 
The projections of the horizontal lines onto the x-axis
represent the time interval each tracer particle
spent in the pile, that is, the transit time of the $i$th tracer
$T(i) = T_{out}(i) - T_{in}(i)$, where $T_{in}(i)$ and 
$T_{out}(i)$ denote the input and output time measured in
units of additions of uncolored grains (1 addition every
$7.7\,s$), respectively.
There is a huge variability in the transit times.
The distribution functions of transit times $P(T,L)$ for all system
sizes are shown in
Fig.\ 3(a). 
A data collapse
for different system sizes $L$ is obtained when plotting
$L^{\beta}\,P(T,L)$ against the rescaled variable
$T/L^{\nu}$ when using $\nu = 1.5 \pm 0.2$ and $\beta = 1.4 \pm 0.2$,
see Fig.\ 3(b). Thus we can write
\begin{equation}
P(T,L) = L^{- \beta}\,F(T/L^{\nu}), \label{fss}
\end{equation}
where $F$ is a scaling function and $\nu$ a critical
exponent expressing how the crossover transit time $T_c$ scales with
system size. 
The scaling function $F$ is of the form
$F(x) = \mbox{constant}$ for $x < 1$ and $F(x) \propto x^{-\alpha}$
for $x > 1$, where $\alpha = 2.4 \pm 0.2$. Since $\alpha > 1$,
it follows
from the form of the scaling function $F$ and the normalization
constraint that $\beta = \nu$. The power-law
tail does not contribute to the mean transit time $\langle T \rangle$
if $\alpha > 2$, and
$\langle T \rangle \propto T_c \propto L^{\nu}$.
In the experiments, the angle of repose was independent of system size.
Thus the average velocity of tracer particles scales like
$\langle V \rangle \propto L/\langle T
\rangle \propto L^{1-\nu} = L^{-0.5 \pm 0.2}$.
It is quiet surprising, that the average velocity is not a constant
but decreases with increasing system size.
The (tracer) particles have information on the system size! This can only
happen if correlations exists throughout the system.
In the statistically stationary state, the steady influx
of particles is balanced with a steady outflux.
During the time $\Delta t$, each particle moves, on average,
a distance $\Delta t \, \langle V \rangle$. Since the feeding rate
was the same for all system sizes, the number of particles
that crossed the outlet in this time interval was constant,
that is, 
$\Delta t \, \langle V \rangle \, {\lambda}_L
= \mbox{constant}$, where the active zone depth
\begin{equation}
{\lambda}_L = \sqrt{\frac{1}{L}\; \sum_{x=1}^{L} \, \langle\, {[ \, h_{x}(t) - \langle h_{x}(t) {\rangle}_t \, ]}^{2} \, {\rangle}_t},
\end{equation}
$h_{x}(t)$ being the height
at position $x$ at time $t$ and $\langle \, \cdot \, {\rangle}_t$ denotes the
temporal average.
This implies that the average velocity
is inversely proportional to the active zone depth,
$\langle V \rangle \propto 1/{\lambda}_L$. An increase in 
${\lambda}_L$ with system size would be consistent with the
experimental finding of a decreasing average velocity with
increasing system size. 

Further insight into the correlations that exist in this dynamical
state can be extracted from the data.
We notice that, once in a while, many tracers dropped out of the
system at the same time since they were part of a large avalanche
that reached the rim of the pile.
The corresponding ``steps'' in the diagrams, like the one shown in Fig.\ 2,
have many different sizes
and are interwoven in a complex way.
The correlation function
\begin{equation}
c(\tau) = \langle I(i,i+\tau) \rangle, \label{corr}
\end{equation} 
where
\begin{equation}
I(i,i+\tau) = \left\{ \begin{array}{ll}
           1 & \mbox{if $\; |\,T_{out}(i)-T_{out}(i+\tau)\,| < \delta t$} \\
           0 & \mbox{otherwise}
          \end{array}
          \right.
\end{equation}
is the indicator function of simultaneously drop out and
$\langle \, \cdot \, \rangle$ denotes
average over all $i$, gives the probability
that tracer $i$ and $i + \tau$ left the system simultaneously. Using
$\delta t = 10$, this quantity is found to behave
as $c(\tau) \propto \exp (-\tau/{\tau}_c(L))$ for small $\tau$,
where $\tau$ is the time difference
in units of 4 minutes and ${\tau}_c(L) = 5.8 \pm 1.3, 8.7 \pm 2,
19 \pm 4,$ and $34 \pm 4$ is the correlation time
for system sizes $L = 20, 39, 79,$ and $113$, respectively.
In Fig.\ 2, we see that ${\tau}_c(L=113) = 34$ corresponds to the
average size of the steps. 
Since the injection rate
was 20 grains/min, the characteristic number of grains
spanned by these correlated sequences of tracers
were $476, 705, 1539,$ and $2754$.
The solid-like motion of domains described in
Ref.\ \cite{nature95} is probably one aspect of this coherent
motion. 
Using ${\lambda}_L \propto \mbox{volume}/L$
and disregarding the smallest system, we obtain
${\lambda}_L \propto L^{0.25 \pm 0.2}$, consistent within
error bars with the result
above.
However, there is significant dispersion.
Grains are being transferred between different
coherently moving areas, see Fig.\ 2.

Inspired by the experiments, we
considered a refined version of a simple $1D$ cellular automaton
studied in Ref. \cite{Frette93}. In a system of size $L$,
an integer variable $h_x$ gives the height
of the pile at site $x$. The local slope
$z_x$ at site $x$ is given by
$z_x = h_x - h_{x+1}$, and we impose $h_{L+1} = 0$.
The addition of a grain at the wall increases the slope by one
at $x = 1$, that is,
$z_1 \rightarrow z_1 + 1.$
We proceed by dropping grains at the wall until the slope $z_1$
exceeds a critical value,
$z_1 > z_{1}^{c}$, then the site topples by transferring
one grain to its neighboring site on the right, $x = 2$. If
$z_x > z_{x}^{c}$, this site
topples in turn according to
\begin{equation}
 \left. \begin{array}{l}
 z_x \rightarrow z_x - 2 \\
 z_{x \pm 1} \rightarrow z_{x \pm 1} + 1
 \end{array}
 \right.
 \label{toppling}
\end{equation}
(unless at the rightmost site where the grains fall off the
pile) generating an avalanche.
During the avalanche, no grains are added to the pile. Thus
the two time scales involved in the dynamic
evolution of the pile are separated. The injection rate of grains is low
compared to the duration of the relaxation processes.
The avalanche stops when the system reaches a stable state
with $z_x \leq z_{x}^{c} \:\: \forall x$ and
grains are added at the wall until a new
avalanche is initiated and so on.
The critical slopes $z_{x}^{c}$ are dynamical variables chosen randomly
to be $1$ or $2$ every time site $x$ has toppled. This is a
simple way to model the changes in the local slopes
observed in the rice pile experiment.
Thus the model differs from the trivial $1D$ Bak, Tang, Wiesenfeld
model where $z_{x}^{c} = 1$ is a constant and where
grains are added on randomly chosen sites \cite{BTW87}.
The randomness in the BTW model is external. In our model,
the randomness is internal and inherent in the dynamics.
Starting with, say, $z_{x} = 0$ and $z_{x}^{c} = 1\:\: \forall x$,
the system reaches a stationary state where the avalanche sizes are
power-law distributed with an exponent of $-1.55 \pm 0.10$ and a cutoff
in the power-law distribution
that scales with system size as $L^{2.25 \pm 0.10}$.

We have measured the transit times (in units of added grains) of the
added particles after the pile has reached the stationary critical state.
Using Eq.\ (\ref{fss}), a reasonable data collapse is obtained with
$\nu = 1.30 \pm 0.10$ and $\beta = 1.35 \pm 0.10$ as displayed in
Fig.\ 4. We find $\alpha = 2.22 \pm 0.10$, that is, the results
agree well with the experimental findings.
The average depth of tracer particles during the transport through the system
as a function of the average transit time is displayed as an inset of
Fig.\ 4, and it shows, that the power-law tail in the
distribution of transit times arises from
tracer particles that become deeply buried in the pile,
while tracer particles that stayed close to the surface of the
pile during the transport through the system mainly contributed to
the constant part.
The average velocity
$\langle V \rangle \propto L/\langle T
\rangle \propto L^{1-\nu}=L^{-0.30 \pm 0.10}$,
in fairly good agreement with the experimental result.
Furthermore, in the model, we are able to measure directly the scaling of the active
zone depth with system size. We find ${\lambda}_L \propto L^{0.25 \pm 0.10}$
in agreement with the relation $\langle V \rangle \propto 1/{\lambda}_L$.
Alternatively, a third measure for the active zone depth is available
from the inset in Fig.\ 4, giving ${\lambda}_L \propto \mbox{depth}
\propto L^{0.3 \pm 0.10}$. Moreover, correlation functions calculated from
Eqs. (3) and (4) give an exponential decay for short times with a crossover
to a decaying power law for large times.
Adding grains at the wall and allowing for dynamical critical slopes,
we have seen similar results for the nonlocal limited model
in Ref. \cite{KNWZ89}. Thus the behavior seems to be universal.

In conclusion, this new direction of research sheds light upon the dynamics
of SOC systems in general and granular systems in particular.
We find that transport in 
a self-organized critical granular medium is characterized
by an average grain velocity that approaches zero when the
system size increases. The dynamics of the system displays
comprehensive correlations. These experimental
findings agree well with the behavior seen in simple
one-dimensional computer models of the self-organized
critical pile.

This work has benefited from discussions with
A. Malthe-S{\o}renssen, P. Meakin, and A. D{\'{\i}}az-Guilera.
K. C. gratefully acknowledges the support by
The Danish Natural Science Research Council, and
NFR, the Research Council of Norway.
A. C. appreciates the hospitality of the Cooperative
Phenomena Group at the University of Oslo, Norway as well as
a grant of the Spanish MEC.
V. F. gratefully acknowledges the support by
NFR, the Research Council of Norway.

%\newpage

%FIG.\ 1
\begin{figure}
\caption{A close up photograph of the rice pile with
size $L = 113$. The plate separation/grain length ratio $d/\delta = 0.79$
and most of the grains are aligned along
the flow direction.
Each tracer particle is uniquely color coded;
thus it is possible to measure individual transit times.
This picture was taken after
$638$ out of $800$ tracer particles had been added.
The region shown contains 5 of a total of
73 tracer particles that were inside the pile at this time.} 
\end{figure}

%FIG.\ 2.
\begin{figure}
%\epsfxsize=2.5truein 
%\hskip 0.15truein\epsffile{fig2.ps} 
%\input{/users/g/alvaro/fortran/ricepile/tex/fig2.tex}
%\input{/mn/fidibus/u1/kimchris/papers/new/tracer/figure/fig2.ps}
\caption{A record of the tracer experiment in a pile of
size $L =113$ where a total number of $800$ tracer
particles were added, one every 4th minute.
The tilted line connects the injection times for all the tracers.
The transit time for each tracer
particle is represented by the length of a horizontal line whose
projection onto the x-axis of the left (right) endpoint
equals the time the particle entered $T_{in}$
(left $T_{out}$) the system. The transit time is measured in
units of the number of injections of uncolored grains (no. additions),
1 addition every $7.7\,s$.
Note the large variation in
the transit times $T = T_{out} - T_{in}$ and that,
repeatedly, many tracers left
the system at the same time. 
The horizontal arrow indicates the time at which the
photograph in Fig.1 was taken. Furthermore, the correlation time
${\tau}_c$ (see text related to Eq.\ ({\protect\ref{corr}})) is indicated.
The inset is the part of the full record marked by dashed lines
and shows that the different sizes
of steps are interwoven.}
\end{figure}

%FIG.\ 3.
\begin{figure}
%\epsfxsize=2.5truein 
%\hskip 0.15truein\epsffile{fig3a.ps} 
%\epsfxsize=2.5truein 
%\hskip 0.15truein\epsffile{fig3b.ps} 
%\input{/users/g/alvaro/fortran/ricepile/tex/fig3a.tex}
%\input{/users/g/alvaro/fortran/ricepile/tex/fig3b.tex}
%\input{/mn/fidibus/u1/kimchris/papers/new/tracer/figure/fig3a.ps}
%\input{/users/g/kim/paper/tracer/fig/fig3b.ps}
\caption{(a) The experimental results
for the normalized distribution of transit times
in piles with sizes $L = 20, 39, 79,$ and $113$.
The data have been averaged over exponentially
increasing bins with base $2$ in order to reduce the fluctuations
in the statistics due to the relatively small number of tracer particles. 
The functions are essentially constant for small
$T$ and have a decaying power-law tail with
a slope of $\alpha = 2.4 \pm 0.2$.
These large transit times correspond to tracer
particles which, during the transport
through the systems, become deeply embedded in the pile.
(b) Disregarding the smallest system and
using Eq.\ ({\protect\ref{fss}}), a reasonable data collapse
of the three largest systems
is obtained with $\nu = 1.5 \pm 0.2$ and $\beta = 1.4 \pm 0.2$.}
\end{figure}

%FIG.\ 4.
\begin{figure}
%\epsfxsize=2.5truein 
%\hskip 0.15truein\epsffile{fig4.ps} 
%\input{/users/g/alvaro/fortran/ricepile/tex/fig4a.tex}
%\input{/users/g/alvaro/fortran/ricepile/tex/fig4b.tex}
%\input{/mn/fidibus/u1/kimchris/papers/new/tracer/figure/fig4a.ps}
%\input{/mn/fidibus/u1/kimchris/papers/new/tracer/figure/fig4b.ps}
\caption{A finite-size scaling plot using Eq.\ ({\protect\ref{fss}})
with $\nu = 1.30 \pm 0.10$ and $\beta = 1.35 \pm 0.10$ of the
normalized distribution of transit times
in the numerical model with system sizes $L = 25, 100, 400,$ and $1600$.
The statistics shown correspond to
$10^7$ tracer particles ($10^6$ for $L=1600$),
and the data have been averaged over exponentially increasing
bins with base $1.1$.
The functions are constant for small
transit times and decay as power laws
with a slope of $\alpha = 2.22 \pm 0.10$.
The inset shows that a data collapse
of the correlation between the transit
times and depth of the tracer particles can be obtained using
a finite-size scaling plot analogous to Eq.\ ({\protect\ref{fss}}), with
${\nu}' = 1.20 \pm 0.10$ and ${\beta}' = -0.30 \pm 0.10$.}
\end{figure}
%\end{multicols}

\begin{references}
%\begin{thebibliography}{99}

\bibitem{BTW87}
{P. Bak, C. Tang, and K. Wiesenfeld,}
{Phys. Rev. Lett. {\bf 59}, 381 (1987);}
{Phys. Rev. A {\bf 38}, 364 (1988).}

\bibitem{nature95}
V. Frette, K. Christensen, A. Malthe-S{\o}renssen, J. Feder, T.
J{\o}ssang, and P. Meakin, {\it Nature} {\bf 379}, 49 (1996);
see also M. Kardar, {\it Nature} {\bf 379}, 22 (1996).

\bibitem{HB87}
{S. Havlin and D. Ben-Avraham,}
{Adv. Phys. {\bf 36}, 695 (1987).}

\bibitem{BG90}
{J.-P. Bouchaud and A. Georges,}
{Phy. Rep. {\bf 195}, 127 (1990).}

\bibitem{HK89}
{T. Hwa and M. Kardar,}
{Phys. Rev. Lett. {\bf 62}, 1813 (1989);}
{Phys. Rev. A {\bf 45}, 7002 (1992).}

\bibitem{CCGS90}
{J. M. Carlson, J. T. Chayes, E. R. Grannan, and G. H. Swindle,}
{Phys. Rev. Lett. {\bf 65}, 2547 (1990);}
{J. M. Carlson, E. R. Grannan, and G. H. Swindle,}
{Phys. Rev. E {\bf 47}, 2366 (1993);}
{J. M. Carlson, E. R. Grannan, C. Singh, and G. H. Swindle,}
{Phys. Rev. E {\bf 48}, 688 (1993).}

\bibitem{BJ92}
{P. B\'{a}ntay and M. J\'{a}nosi,}
{Physica A {\bf 185}, 11 (1992).}

\bibitem{Frette93}
{V. Frette,}
{Phys. Rev. Lett. {\bf 70}, 2762 (1993).}
 
\bibitem{KNWZ89}
{L. P. Kadanoff, S. R. Nagel, L. Wu, and S-m. Zhou,}
{Phys. Rev. A {\bf 39}, 6524 (1989).}

%\end{thebibliography}
\end{references}
\end{document}